\documentclass[twocolumn,apj]{emulateapj}
\usepackage{apjfonts}
\epsscale{1.18}
\newcommand{\rotate}{}

\shorttitle{Mergers in Double-Peaked [O\,{\sc iii}] AGNs} 
\shortauthors{Fu et al.}
\journalinfo{to appear in the Astrophys. J.}
\submitted{Received 2010 September 3; Accepted 2011 March 21}

\newcommand{\kms}{{km s$^{-1}$}}
\newcommand{\msun}{$M_{\odot}$}
\newcommand{\lsun}{$L_{\odot}$}
\newcommand{\msigma}{$M_{\rm BH}$$-$$\sigma_\star$}
\newcommand{\sbulge}{$\sigma_\star$}
\newcommand{\nd}{\nodata}
\newcommand{\OIII}{[O\,{\sc iii}]}

\begin{document}

\title{Mergers in Double-Peaked [O\,{\sc iii}] Active Galactic Nuclei\altaffilmark{*}}

\altaffiltext{*}{Some of the data presented herein were obtained at the W.M. Keck Observatory, which is operated as a scientific partnership among the California Institute of Technology, the University of California and the National Aeronautics and Space Administration. The Observatory was made possible by the generous financial support of the W.M. Keck Foundation.}

\author{Hai Fu\altaffilmark{1}, 
Adam D. Myers\altaffilmark{2,3},
S. G. Djorgovski\altaffilmark{1}, and
Lin Yan\altaffilmark{4}
}
\altaffiltext{1}{Astronomy Department, California Institute of Technology, MS 249$-$17, Pasadena, CA 91125, USA; fu@astro.caltech.edu, george@astro.caltech.edu} 
\altaffiltext{2}{Department of Astronomy, University of Illinois at Urbana-Champaign, Urbana, IL 61801, USA; admyers@astro.illinois.edu}
\altaffiltext{3}{Max-Planck-Institut f\"ur Astronomie, K\"onigstuhl 17, D-69117 Heidelberg, Germany}
\altaffiltext{4}{Spitzer Science Center, California Institute of Technology, MS 220$-$06, Pasadena, CA 91125, USA; lyan@ipac.caltech.edu}  

\begin{abstract} 

As a natural consequence of galaxy mergers, binary active galactic nuclei (AGNs) should be commonplace. Nevertheless, observational confirmations are rare, especially for binaries with separations less than ten kpc. Such a system may show two sets of narrow emission lines in a single spectrum owing to the orbital motion of the binary. We have obtained high-resolution near-infrared images of 50 double-peaked \OIII\,$\lambda$5007 AGNs with the Keck\,II laser guide star adaptive optics system. The Sloan Digital Sky Survey sample is compiled from the literature and consists of 17 type-1 AGNs between $0.18 < z < 0.56$ and 33 type-2 AGNs between $0.03 < z < 0.24$. The new images reveal eight type-1 and eight type-2 sources that are apparently undergoing mergers. These are strong candidates of kpc-scale binary AGNs, because they show multiple components separated between 0.6 and 12 kpc and often disturbed morphologies. Because most of the type-1s are at higher redshifts than the type-2s, the higher merger fraction of type-1s ($47\pm20$\%) compared to that of type-2s ($24\pm10$\%) can be attributed to the general evolution of galaxy merger fraction with redshift. Furthermore, we show that AGN mergers are outliers of the \msigma\ relation because of over-estimated stellar velocity dispersions, illustrating the importance of removing mergers from the samples defining the \msigma\ relations. Finally, we find that the emission-line properties are indistinguishable for spatially resolved and unresolved sources, emphasizing that scenarios involving a single AGN can produce the same double-peaked line profiles and they account for at least 70\% of the double-peaked \OIII\ AGNs. 

\end{abstract}

\keywords{galaxies: active --- galaxies: formation --- galaxies: interactions --- galaxies: nuclei --- quasars: emission lines}

\section{Introduction} \label{sec:introduction}

Mergers play an essential role in the assembly and evolution of galaxies, because structures form hierarchically in the cold dark matter dominated Universe. Deep cosmological surveys have shown that roughly 10\% of $L^\star$ galaxies are involved in major mergers at $z$ = 1, and the merger fraction varies with galaxy mass and redshift \citep[see][and references therein]{Hopkins10b}. As supermassive black holes (SMBHs) reside in the centers of most, if not all, galaxies \citep[e.g.,][]{Kormendy95,Richstone98}, the standard cosmological paradigm predicts the formation of binary SMBHs from galaxy mergers. 

Identifying binary SMBHs is of a fundamental interest. The population of binary SMBHs depends critically on the efficiency of their coalescence (the so-called {\it Final Parsec Problem}; \citealt{Begelman80}); moreover, whether the binary stalls over the final few parsecs \citep[e.g.,][]{Milosavljevic01}, coalesces \citep[e.g.,][]{Escala04}, or recoils \citep[e.g.,][]{Madau04} has key implications for future gravitational wave detection and the demography of SMBHs. 

\begin{figure*}[!t]
\plotone{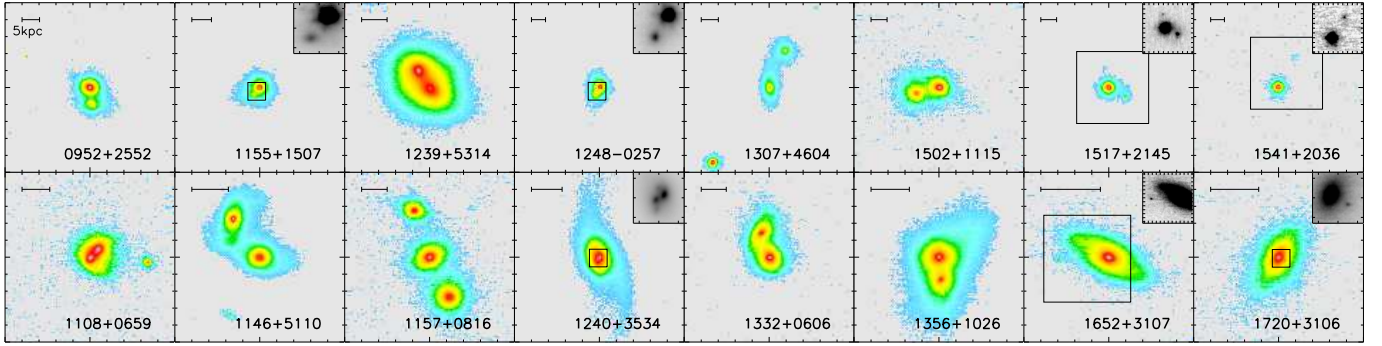}
\caption{Keck\,II/LGSAO images of the double-peaked AGNs with multiple components (\emph{top}: type-1s, \emph{bottom}: type-2s). The pseudo-color images are displayed in $asinh$ (the inverse hyperbolic sine function) scales to bring up low surface brightness features. North is up and east is to the left (10\arcsec\ stamps with 1\arcsec\ tickmarks). The images of 0952+2552 and 1240+3534 are in $H$-band, and the rest are in $K'$-band. The insets show the components in linear scales for extremely compact systems ($\Delta\theta <$ 0\farcs6) or those of large contrasts ($\Delta K' > 2.5$). The open boxes delineate the regions covered by the insets. The tickmarks in the insets are spaced in 0\farcs4. The scale bar indicates a transverse separation of 5~kpc. Confirmation of three of the type-2 binary systems (1108+0659, 1146+5110, and 1332+0606) with seeing-limited imaging and long-slit spectroscopy previously appeared in \citet{Liu10b}. \citet{Smith10} also noticed the unusual morphologies of 1157+0816 and 1307+4604 in the SDSS images.
\label{fig:NIRC2}} 
\end{figure*}

\begin{figure*}[!t]
\plotone{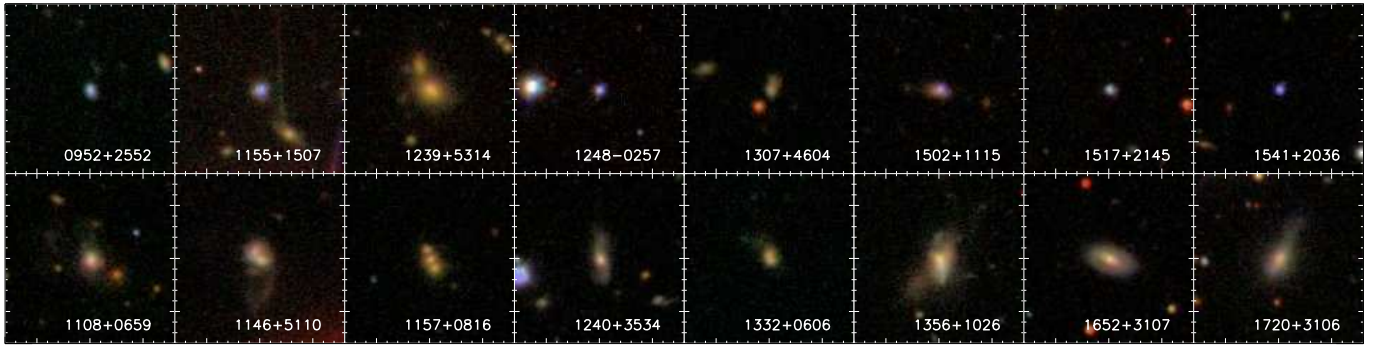}
\caption{SDSS $gri$ color-composite images. North is up and east is to the left (48\arcsec\ stamps with 3\arcsec\ tickmarks). Arranged in the same sequence as in Fig.~\ref{fig:NIRC2}.
\label{fig:SDSS}} 
\end{figure*}

Before the galaxies coalesce and their SMBHs start to orbit each other, a fraction of the mergers may appear as binary active galactic nuclei (AGNs). Gravitational interactions can drive gas from the galactic outskirts into the nuclei of the merging galaxies, as shown in simulations \citep[e.g.][]{Barnes96} and observations \citep[e.g.,][]{Fu07a, Kewley10}. The rapid gas concentration may fuel intensive black hole accretion. Although AGN pairs with $>$10~kpc transverse separations comprise $\sim$0.1\% of the AGN population \citep[e.g.][]{Hennawi06,Hennawi10,Myers07,Myers08,Green10}, there are surprisingly few known kpc- to parsec-scale binaries \citep[e.g.,][]{Komossa03,Valtonen08,Rodriguez09}, owing in large to the observational difficulty of spatially resolving such binaries.

Spatially unresolved binaries may be spectroscopically resolved, thanks to the orbital motion of the binaries. AGNs with pairs of broad Balmer emission lines were believed to be strong candidates of parsec-scale binary SMBHs \citep[e.g.,][]{Gaskell83,Gaskell96}. But such systems could be confused with single AGNs with unusual broad line regions or thin-plus-thick accretion disks \citep[e.g.,][]{Eracleous97,Eracleous03}; and it is difficult to distinguish these scenarios because they all involve very small spatial scales ($\lesssim$ 1 pc). The discovery of the double-peaked broad-line AGN SDSS~J153636.221+044127.0 \citep{Boroson09} prompted \citet{Gaskell10} to reiterate the ``smoking gun" criteria for the confirmation of a binary SMBH. First, the two broad emission peaks should shift on timescales corresponding to an orbit, much like a spectroscopic binary star. Second, because of the known variability of broad line regions, variations in the line fluxes of the two emission peaks should be independent. These criteria may be difficult to met for SDSS~J153636.221+044127.0, which displays no spectral variability in the observed frame in a year \citep{Chornock10}. To this date, no \emph{unambiguous} parsec-scale binary SMBHs have been found with this method.

Similarly, systematic searches for kpc-scale binary AGNs started from identifying objects with two sets of narrow \OIII\,$\lambda$5007 emission lines in large spectroscopic surveys, such as the DEEP2 Galaxy Redshift Survey \citep{Gerke07,Comerford09a} and the Sloan Digital Sky Survey \citep[SDSS;][]{Xu09,Wang09,Liu10a,Smith10}. If double-peaked \OIII\ emission is indicative of a binary AGN, such emission most likely represents systems with transverse separations of $\sim$100 pc to $\sim$10 kpcs \citep{Liu10b}, because each fiber of the SDSS spectrograph covers only 3\arcsec\ on the sky. But double-peaked emission-line profile may also arise from other mechanisms, such as peculiar narrow line regions, extended emission-line nebulae \citep[e.g.,][]{Fu09a}, or jet-cloud interactions \citep[e.g.,][]{Stockton07,Rosario10}.

In this paper, we present high-resolution near-infrared imaging of double-peaked \OIII\ AGNs. This program is designed to differentiate between mergers and the contaminating scenarios involving single galaxies, as an attempt to identify a statistical sample of kpc-scale binary AGNs. 
Note that spatially resolved spectroscopy is required to confirm that a merger is a binary AGN, because the double-peaked emission lines could still arise from a single galaxy in a merger or extended emission-line nebulae.

Throughout we adopt the concordance $\Lambda$CDM cosmology with $\Omega_{\rm m}=0.3$, $\Omega_\Lambda=0.7$, and $H_0$ = 70 km~s$^{-1}$~Mpc$^{-1}$.

\section{Observations and Data Reduction}

Keck\,II laser guide-star adaptive-optics \citep[LGSAO;][]{Wizinowich06} observations require a bright ($R < 17$) star within $\sim$60\arcsec\ of the source for tip-tilt corrections. We selected 125 objects that are suitable for LGSAO observations from the 271 unique SDSS double-peaked \OIII\ AGNs \citep[][we only included ``good" objects]{Wang09,Liu10a,Smith10}, resulting in 31 type-1 (broad-line) AGNs and 94 type-2 (narrow-line) AGNs between $0.02 < z < 0.69$. 

We obtained $H$-band images for three sources (0400-0652, 0952+2552, and 1240+3534) with the OSIRIS imager \citep{Larkin06} at 20 mas pixel$^{-1}$ on 2010 March 6 and 7 UT, and $K'$-band images for another 47 sources with NIRC2 at 40 mas pixel$^{-1}$ on 2010 June 3 and 4 UT. 
We imaged only three sources in March because spectroscopy was the main purpose of that run. The $K'$-band filter was used for most sources because (1) the AO system delivers the best image quality in $K'$-band, and (2) the contrast between the central AGN and the host galaxy is decreased because AGNs tend to be bluer than galaxies. On the other hand, we chose $H$-band for OSIRIS because of its elevated background in $K$-band (OSIRIS User's Manual).

Targets, which were randomly selected based on time of observation, consisted of 17 type-1s and 33 type-2s (Table~\ref{tab:sample}). A typical imaging sequence consisted of 5 to 9 frames of 64 s exposures, dithered within boxes of 4\arcsec\ to 7\arcsec. Conditions were photometric and the seeing was about 0\farcs7 for the OSIRIS nights and 0\farcs7/0\farcs5 for the first/second NIRC2 night. 

Data were processed using an iterative IRAF reduction pipeline \citep{Stockton98}. Before individual frames were combined, we corrected the NIRC2 geometric distortions using the solution of P. B. Cameron\footnote{http://www2.keck.hawaii.edu/inst/nirc2/forReDoc/post\_observing/dewarp/}. The PSF (point-spread function) FWHMs range from 0\farcs065 to 0\farcs130 with a mean of $\sim$0\farcs1, as measured from stellar sources in the coadded images. For our tip-tilt configurations and seeing conditions, the $K'$-band Strehl ratio is $\sim$0.2 \citep{Dam06}.

\section{Analyses and Results}

\subsection{Mergers among Double-Peaked AGNs}

\begin{figure}
\plotone{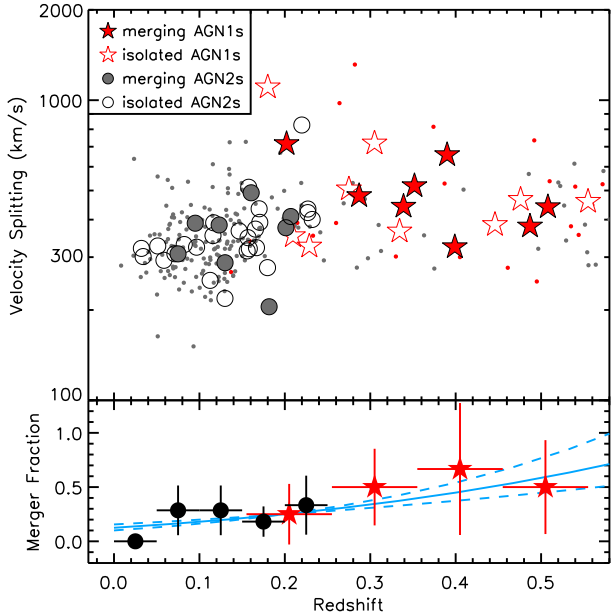}
\caption{{\it Top}: \OIII\ velocity splitting vs. redshift. Double-peaked AGNs with LGSAO images are large red stars (type-1s) and black circles (type-2s). Filled (open) symbols represent merging (isolated) objects. The average statistical uncertainty of the velocity splitting measurements is $\sim$16 \kms, comparable to the size of the symbols at the lower bound ($\sim$ 200 \kms). The AGNs without LGSAO images are shown as small red (type-1s) and grey (type-2s) filled circles. {\it Bottom}: Merger fraction vs. redshift for type-1s (red stars) and type-2s (black circles). Horizontal error bars show the bin size, and vertical error bars are Poisson errors. The resolved fraction increases with redshift. Solid and dashed blue curves show the merger fraction evolves as $(1+z)^{3.8\pm1.2}$ \citep{Kampczyk07} normalized at 25\% at $z = 0.2$.
\label{fig:resolve}} 
\end{figure}

Sixteen of the 50 double-peaked AGNs appear to be undergoing mergers with multiple components within 3\arcsec\ (Fig.~\ref{fig:NIRC2}; Table~\ref{tab:sample}). Note that spectroscopic confirmation of three of the type-2 systems (1108+0659, 1146+5110, and 1332+0606) previously appeared in \citet{Liu10b}. For most of the eight type-2 systems, signs of interactions such as tidal tails and asymmetries are evident in the Keck and SDSS images (bottom rows of Figs.~\ref{fig:NIRC2} \& \ref{fig:SDSS}). Signs of interactions are less clear for the eight type-1 systems (top rows of Figs.~\ref{fig:NIRC2} \& \ref{fig:SDSS}). Nevertheless, at least five of the type-1 systems are definitely mergers, as we measured small (hundreds \kms) redshift differences between the components with spatially resolved spectroscopy (H. Fu et al. 2011, in preparation).

We used GALFIT \citep{Peng10} to model the merging systems. PSFs were taken either from stellar sources inside the same image or from time-adjacent images with similar tip-tilt configurations. We matched the combined magnitudes of the merging components to their 2MASS magnitudes because the components are mostly unresolved in 2MASS (except for 1157+0816). We transformed filters from $K_S$ to $K'$ following John Carpenter's calibrations\footnote{http://www.astro.caltech.edu/$\sim$jmc/2mass/v3/transformations/} and the Mauna Kea Observatory $K'$ definition \citep{Wainscoat92}. The result are summarized in Table~\ref{tab:sample}. 

For 11 of the 16 objects, the $K'$ magnitudes of the merging components differ by less than 1.2 magnitudes, satisfying the definition of a major merger (mass ratio $\mu > 1/3$ given a constant mass-to-light ratio). Note that the primary components of many type-1 systems are brightened by the central point sources. For that reason, 0952+2552 is likely a major merger even though it shows a magnitude difference of 1.6 mag. 

The 16 multiple component sources split evenly between type-1s and type-2s. Because we observed a total of 17 type-1s and 33 type-2s, the merger fraction of type-1s ($47\pm20$\%) is greater than that of type-2s ($24\pm10$\%). This is surprising because it should be more difficult to detect extended companion galaxies around type-1s than type-2s owing to the bright nuclei of the former.  

In light of the rapid evolution of galaxy merger fraction at $z < 2.5$ \citep[see the model of][and references therein for observational constraints]{Hopkins10b}, the higher merger fraction in type-1s is likely a result of the distinct redshift distributions of type-1 and type-2 AGNs (Figure~\ref{fig:resolve}), as opposed to an intrinsic difference between the two groups. 
Ideally, we would want to compare the merger fraction of AGNs with our data. In practice, the former is not well determined because of the lack of high-resolution images and the rarity of AGNs in deep cosmological surveys. Nevertheless, based on the 2 deg$^2$ COSMOS {\it Hubble} Space Telescope ACS images \citep{Scoville07b}, \citet{Cisternas11} found that there is no significant difference in the distortion fractions between active and inactive galaxies. 
Therefore, as a surrogate, in the bottom panel of Fig.~\ref{fig:resolve} we compare our data with the result of \citet{Kampczyk07} ($f_{\rm merger} \propto (1+z)^{3.8\pm1.2}$), who determined the evolution of galaxy merger fraction using COSMOS ACS images after statistically removing spurious mergers by simulating the appearance of real low-redshift SDSS galaxies at $z = 0.7$. Our data are consistent with the evolution of the major merger fraction in typical $L^\star$ galaxies, although a larger sample is clearly needed to improve the statistics. Note that by selection mergers are over an order-of-magnitude more common in double-peaked AGNs than in $L^\star$ galaxies in the same redshift range \citep[see the compilation of][]{Hopkins10b}. 

\subsection{The Impact of Mergers on the \msigma\ relation} \label{sec:msigma}

\begin{figure}
\plotone{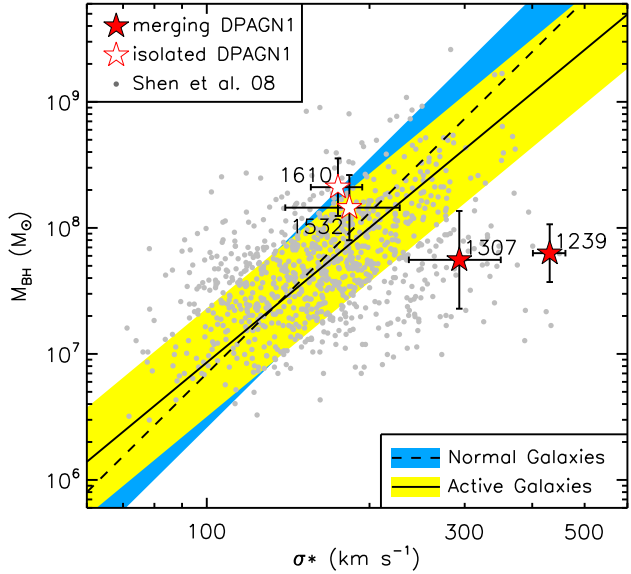}
\caption{Black hole mass vs. stellar velocity dispersion. AGNs with AO-corrected images are red stars with error bars (open symbol: isolated galaxies, filled symbol: mergers). Their RA's are labeled. The SDSS AGN sample of \citet{JJShen08} are shown as grey circles. Measurements were made with single-epoch SDSS spectra. The solid line is the best-fit \msigma\ relation from local Seyfert 1s with reverberation mapping black hole masses \citep{Woo10}, and the dashed line is the best-fit relation from quiescent galaxies \citep{Gultekin09}; the color shaded areas delimit the $\sim0.43$ dex intrinsic scatter.  
\label{fig:m_sigma}} 
\end{figure}

The \msigma\ relation suggests that black hole masses correlate tightly with the velocity dispersions of stellar bulges (\sbulge). Although the \msigma\ relation is considered fundamental, it remains under debate whether active galaxies and inactive galaxies lie on the same \msigma\ relation. The small sample of reverberation-mapped Seyfert 1 galaxies suggest that they do \citep[e.g.,][]{Woo10,Graham11}, while the larger sample of SDSS AGNs suggest not \citep{Greene06,JJShen08}. The latter groups used single-epoch spectra to estimate black hole masses and found that the \msigma\ relation of AGNs is shallower than that of quiescent galaxies. One of the main differences between reverberation and single-epoch black hole masses is that the former uses the the width of the \emph{variable} part of the broad emission line measured from root mean square (rms) spectra \citep[][]{Peterson98}. Single-epoch emission line widths show $\sim$25\% variation when compared with the rms line widths \citep[][]{Vestergaard02}. As the black hole mass depends sensitively on the line width, this variation may lead to the discrepancy in the \msigma\ relations. 

On the other hand, AGNs involved in galaxy mergers can easily deviate from the \msigma\ relation because the velocity dispersions are broadened by the relative motions of the components. To test this idea, we measured single-epoch black hole masses and velocity dispersions for the broad-line AGNs in our sample. Because velocity dispersion measurements require a significant host galaxy contribution in the optical spectra, we were only able to measure velocity dispersions in four type-1 AGNs in our sample (1239+5314, 1307+4604, 1532+4203, and 1610+1308). 

We measured single-epoch black hole masses with the SDSS spectra and the calibration of \citet{Vestergaard06}. We measured FWHMs of the broad H$\beta$ or H$\alpha$ lines. We did not decompose the broad lines for the kinematic components seen in the double-peaked narrow lines, because (1) the decomposition result would be highly uncertain, and (2) the companions of the type-1 AGNs are unlikely to be type-1 AGNs since most of them look extended from the LGSAO images. H$\alpha$ line FWHMs were converted to H$\beta$ FWHMs using the relation given by Eq. [2] of \citet{JJShen08}. We corrected the rest-frame monochromatic continuum luminosity at 5100 \AA, $L_{5100}$, for Galactic extinction using the \citet{Cardelli89} law and for host galaxy contamination using the empirical fit of \citet[][]{Shen10}. $L_{5100}$ scaled from \OIII\,$\lambda5007$ luminosities gave similar results. For the objects in common, our black hole masses agree with those from \citet{JJShen08}\footnote{One should add $\sim0.1$ dex to the black hole masses from \citet{JJShen08} because SDSS DR3 spectra were not corrected for fiber flux loss.} and \citet{Shen10} within 0.1 dex. Besides measurement errors in the FWHMs and $L_{5100}$, we also included 0.1 dex uncertainty in the virial coefficient \citep{Woo10} and 0.2 dex uncertainty in the size-luminosity relation of the broad line regions \citep{Kaspi05} in the 1-$\sigma$ uncertainty of black hole masses.

We measured stellar velocity dispersions using the same SDSS spectra with the IDL routine {\it vdispfit}\footnote{http://spectro.princeton.edu/idlspec2d\_install.html}, which determines the velocity dispersion and its error by fitting the galaxy spectrum with a number of Gaussian-broadened stellar template spectra. We masked out emission lines, telluric absorption, and artifacts from sky subtraction in the fitting. Following \citet{JJShen08}, the velocity dispersions were corrected for the 3\arcsec\ fiber size, which changes \sbulge\ by less than 10\% at $z < 0.6$. For the object in common (1239+5314), our measurement ($431\pm30$ \kms) agrees with that of \citet{JJShen08} ($449\pm30$ \kms). 

The results are shown in Figure~\ref{fig:m_sigma}. The two double-peaked AGNs in mergers fall significantly below the \msigma\ relations of inactive galaxies and reverberation-mapped Seyfert 1s, while the isolated double-peaked AGNs lie on the relations within the 1-$\sigma$ scatter. This result highlights the importance of identifying mergers with high-resolution imaging before comparing the \msigma\ relation of AGNs and normal galaxies. On the other hand, mergers may also affect the \msigma\ relation of quiescent galaxies, as mergers constitute almost half of the early-type galaxies with velocity dispersions greater than 350 \kms\ \citep[][]{Bernardi08}.

\citet{JJShen08} found that AGNs with larger Eddington ratios ($L_{\rm bol}/L_{\rm Edd}$) tend to have larger stellar velocity dispersions than those with lower Eddington ratios at a given black hole mass. In combination with our result, it seems to imply that mergers show larger Eddington ratios, or more efficient black hole accretion. Here we have assumed that the AGNs that show excess in stellar velocity dispersions are all mergers. Future high-resolution imaging of the AGNs in \citeauthor{JJShen08}'s sample will be able to test this hypothesis.

\subsection{Isolated Double-Peaked AGNs}\label{sec:id_bAGN}

\begin{figure}
\plotone{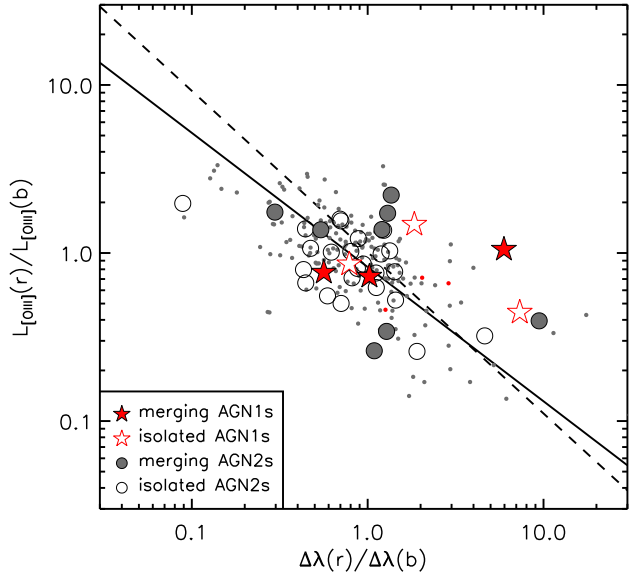}
\caption{Luminosity ratio of the red and blue \OIII\ components vs. ratio of their wavelength offsets relative to the systemic redshifts. Symbols are the same as in Fig.~\ref{fig:resolve} and are labeled. The solid line is a least squares bisector fit of the entire sample. The dashed line is the best-fit from \citet{Wang09}, whose sample is restricted to type-2 AGNs at $z < 0.15$. 
\label{fig:virial}} 
\end{figure}

Our images have shown that $\sim$70\% of double-peaked AGNs are not mergers. To increase the success rate of binary identification, it is natural to investigate whether the merging double-peaked AGNs look different from the isolated double-peaked AGNs in terms of their \OIII\ profiles. Because our sample was compiled from three different groups, we re-modeled the emission lines to homogenize the measurements. We decomposed the \OIII\,$\lambda5007$ line with either two Gaussians or Lorentzian profiles, whichever gives the lower $\chi^2$. We found that the merging and isolated double-peaked AGNs show indistinguishable \OIII\ profiles and velocity splittings (Fig.~\ref{fig:resolve}).

\citet{Wang09} found an anti-correlation between flux ratio and velocity shift ratio of the \OIII\ components, which they interpreted as an indication of the orbital motion of binary AGNs. Such an anti-correlation is expected if (1) the \OIII\ luminosities correlate with the black hole masses, and (2) the black hole masses correlates with the bulge masses (the Magorrian relation). For the former statement to hold, they had to assume identical bolometric corrections and Eddington ratios for both components in each binary system. For the subsample where systemic redshifts could be measured from stellar absorption lines (compiled from \citealt{Wang09,Liu10a,Smith10}), we found that both merging and isolated double-peaked AGNs fall on the same anti-correlation with similar dispersions (Fig.~\ref{fig:virial}). Because the isolated double-peaked AGNs cannot be kpc-scale binary AGNs, these results show that other mechanisms, such as jet-driven outflows, extended emission-line nebulae, and peculiar kinematics in the narrow line regions, can produce the same double-peaked line profiles {\em and} the apparent virial relation of \citet{Wang09}. 

Jet-cloud interactions are an important mechanism in producing double-peaked \OIII\ lines \citep[e.g.,][]{Stockton07,Rosario10}. Although the radio-detected fraction of type-2 double-peaked AGNs is similar to that of the general type-2 AGN population \citep[$\sim$35\%;][]{Liu10a}, \citet{Smith10} found that radio sources are three times over-represented in type-1 double-peaked AGNs than in the general type-1 AGN population. Are radio-detected type-1 AGNs also over-represented in those that are undergoing mergers? To include objects with complex radio morphologies, we obtained 1.4~GHz radio continuum images of the double-peaked AGNs from the FIRST survey \citep{Becker95}. For broad-line double-peaked AGNs as a whole, the radio detected fraction is 34\% (14/41). The fraction is slightly higher (47\%) for the 17 broad-line double-peaked AGNs with AO-corrected images. For the 8 sources that appear to be mergers, the fraction is 38\% (3/8). Although radio detections are less common in mergers (38\%) compared to isolated AGNs (56\%), the radio-detected fraction remains three times (after correcting for the bias in the AO sample) higher than that of the overall broad-line AGN population \citep[$\sim$10\% at $0.2<z<0.5$;][]{Schneider07}. A similar fraction of merging type-2s are radio detected (38\%, 3/8). Therefore, jet-cloud interactions should also be an important mechanism in producing double-peaked emission-lines in AGN mergers.

\section{Discussion}

In recent years, there has been a flurry of observational and theoretical research on binary SMBHs/AGNs. Although a few binary SMBH/AGN candidates have been known since late nineties \citep{Kochanek99}, the availability of large spectroscopic surveys such as SDSS has rejuvenated this subject and enabled systematic searches for large samples of binary candidates based on spectroscopic signatures indicating potentially two AGN components orbiting each other. The recent progress is driven by the predictions of the standard hierarchical galaxy formation paradigm, the lack of an observational understanding of the formation and evolution of binary SMBHs, and the requirement for making predictions of gravitational wave signals potentially detectable by LIGO and the future laser interferometry mission (LISA). 

With several published catalogs of binary AGN candidates, the remaining critical issue is to determine the characteristics of a true binary AGN and the fraction of genuine binary AGNs among the candidates. We carried out a high-resolution imaging survey of double-peaked SDSS AGNs. The experiment was designed to differentiate mergers from contaminating sources involving single AGNs. In 50 sources we found 16 that are apparently undergoing mergers. Our analyses demonstrate clearly that a large percentage ($>$70\%) of these binary AGN candidates are isolated AGNs. Their double peaked \OIII\ line profiles could be explained by several other processes, as discussed in \S~\ref{sec:id_bAGN}. 

Because only $\sim$1\% of the SDSS AGNs show double-peaked line profiles, the $\sim$30\% merger fraction thus indicates a kpc-scale binary AGN fraction\footnote{i.e., the fraction of binary AGNs among all AGNs.} of $<$0.3\%, which is comparable to the fraction of AGN pairs with $>$10 kpc transverse separations ($\sim$0.1\%). There are several factors that make this binary fraction just an order of magnitude estimate. Firstly, it is possible that one of the merging components is actually inactive and the double-peaked line profile is produced by a single galaxy. In a forthcoming paper, we will present spatially resolved spectroscopy of these AGN mergers to securely identify binary AGNs. Secondly, binaries with small line-of-sight velocity separations ($\lesssim$ 200 \kms; see Fig.~\ref{fig:NIRC2}) would not show double-peaked emission lines in the SDSS spectra. These spectroscopically unresolved binaries may comprise bulk of the kpc-scale binary AGN population \citep[e.g.,][]{Comerford09a}. Because $\sim$99\% of the SDSS AGNs do not show double-peaked emission lines, future high-resolution studies of single-peaked AGNs will provide better constraints on the binary AGN fraction. 

Assuming that (1) galaxy mergers do \emph{not} enhance AGN activities and (2) the merger fractions are similar for active and inactive galaxies \citep{Cisternas11}, we would expect a binary AGN fraction of $\sim$0.05\% at $z \sim 0.5$. We calculated the number by multiplying the average AGN duty cycle ($\sim$1\%) and the AGN merger fraction ($\sim$5\% for $L^{\star}$ galaxies at $z \sim 0.5$; \citealt{Hopkins10b}). Therefore, a well-determined binary AGN fraction much higher than the expected value may provide the long-sought observational evidence for merger-induced AGN activities.

\acknowledgments
We thank the referee for a careful reading of the manuscript and for his/her comments that helped to improve the paper. ADM acknowledges partial support from NASA (grants NNX08AJ28G and GO9-0114). SGD was partially supported by NSF grant AST-0909182 and the Ajax Foundation. The authors wish to recognize and acknowledge the very significant cultural role and reverence that the summit of Mauna Kea has always had within the indigenous Hawaiian community.  We are most fortunate to have the opportunity to conduct observations from this mountain.

\facility{{\it Facilities}: Keck:II (LGSAO/NIRC2, LGSAO/OSIRIS), Sloan, FLWO:2MASS, CTIO:2MASS}

\newpage

\begin{thebibliography}
\expandafter\ifx\csname natexlab\endcsname\relax\def\natexlab#1{#1}\fi

\bibitem[{Barnes \& Hernquist(1996)}]{Barnes96}
Barnes, J.~E., \& Hernquist, L. 1996, \apj, 471, 115

\bibitem[{Becker {et~al.}(1995)Becker, White, \& Helfand}]{Becker95}
Becker, R.~H., White, R.~L., \& Helfand, D.~J. 1995, \apj, 450, 559

\bibitem[{Begelman {et~al.}(1980)Begelman, Blandford, \& Rees}]{Begelman80}
Begelman, M.~C., Blandford, R.~D., \& Rees, M.~J. 1980, Nature, 287, 307

\bibitem[{Bernardi {et~al.}(2008)Bernardi, Hyde, Fritz, Sheth, Gebhardt, \&
  Nichol}]{Bernardi08}
Bernardi, M., Hyde, J.~B., Fritz, A., Sheth, R.~K., Gebhardt, K., \& Nichol,
  R.~C. 2008, \mnras, 391, 1191

\bibitem[{Boroson \& Lauer(2009)}]{Boroson09}
Boroson, T.~A., \& Lauer, T.~R. 2009, Nature, 458, 53

\bibitem[{Cardelli {et~al.}(1989)Cardelli, Clayton, \& Mathis}]{Cardelli89}
Cardelli, J.~A., Clayton, G.~C., \& Mathis, J.~S. 1989, \apj, 345, 245

\bibitem[{Chornock {et~al.}(2010)Chornock, Bloom, Cenko, Filippenko, Silverman,
  Hicks, Lawrence, Mendez, Rafelski, \& Wolfe}]{Chornock10}
Chornock, R., {et~al.} 2010, \apj, 709, L39

\bibitem[{Cisternas {et~al.}(2011)Cisternas, Jahnke, Inskip, Kartaltepe,
  Koekemoer, Lisker, Robaina, Scodeggio, Sheth, Trump, Andrae, Miyaji, Lusso,
  Brusa, Capak, Cappelluti, Civano, Ilbert, Impey, Leauthaud, Lilly, Salvato,
  Scoville, \& Taniguchi}]{Cisternas11}
Cisternas, M., {et~al.} 2011, \apj, 726, 57

\bibitem[{Comerford {et~al.}(2009)Comerford, Gerke, Newman, Davis, Yan, Cooper,
  Faber, Koo, Coil, Rosario, \& Dutton}]{Comerford09a}
Comerford, J.~M., {et~al.} 2009, \apj, 698, 956

\bibitem[{Eracleous \& Halpern(2003)}]{Eracleous03}
Eracleous, M., \& Halpern, J.~P. 2003, \apj, 599, 886

\bibitem[{Eracleous {et~al.}(1997)Eracleous, Halpern, Gilbert, Newman, \&
  Filippenko}]{Eracleous97}
Eracleous, M., Halpern, J.~P., Gilbert, A.~M., Newman, J.~A., \& Filippenko,
  A.~V. 1997, \apj, 490, 216

\bibitem[{Escala {et~al.}(2004)Escala, Larson, Coppi, \& Mardones}]{Escala04}
Escala, A., Larson, R.~B., Coppi, P.~S., \& Mardones, D. 2004, \apj, 607, 765

\bibitem[{Fu \& Stockton(2007)}]{Fu07a}
Fu, H., \& Stockton, A. 2007, \apj, 666, 794

\bibitem[{Fu \& Stockton(2009)}]{Fu09a}
---. 2009, \apj, 690, 953

\bibitem[{Gaskell(1983)}]{Gaskell83}
Gaskell, C.~M. 1983, in Liege International Astrophysical Colloquia, Vol.~24,
  473--477

\bibitem[{Gaskell(1996)}]{Gaskell96}
Gaskell, C.~M. 1996, \apj, 464, L107

\bibitem[{Gaskell(2010)}]{Gaskell10}
---. 2010, Nature, 463, 1

\bibitem[{Gerke {et~al.}(2007)Gerke, Newman, Lotz, Yan, Barmby, Coil,
  Conselice, Ivison, Lin, Koo, Nandra, Salim, Small, Weiner, Cooper, Davis,
  Faber, \& Guhathakurta}]{Gerke07}
Gerke, B.~F., {et~al.} 2007, \apj, 660, L23

\bibitem[{Graham {et~al.}(2011)Graham, Onken, Athanassoula, \&
  Combes}]{Graham11}
Graham, A.~W., Onken, C.~A., Athanassoula, E., \& Combes, F. 2011, \mnras, 48

\bibitem[{Green {et~al.}(2010)Green, Myers, Barkhouse, Mulchaey, Bennert, Cox,
  \& Aldcroft}]{Green10}
Green, P.~J., Myers, A.~D., Barkhouse, W.~A., Mulchaey, J.~S., Bennert, V.~N.,
  Cox, T.~J., \& Aldcroft, T.~L. 2010, \apj, 710, 1578

\bibitem[{Greene \& Ho(2006)}]{Greene06}
Greene, J.~E., \& Ho, L.~C. 2006, \apj, 641, L21

\bibitem[{G{\"u}ltekin {et~al.}(2009)G{\"u}ltekin, Richstone, Gebhardt, Lauer,
  Tremaine, Aller, Bender, Dressler, Faber, Filippenko, Green, Ho, Kormendy,
  Magorrian, Pinkney, \& Siopis}]{Gultekin09}
G{\"u}ltekin, K., {et~al.} 2009, \apj, 698, 198

\bibitem[{Hennawi {et~al.}(2010)Hennawi, Myers, Shen, Strauss, Djorgovski, Fan,
  Glikman, Mahabal, Martin, Richards, Schneider, \& Shankar}]{Hennawi10}
Hennawi, J.~F., {et~al.} 2010, \apj, 719, 1672

\bibitem[{Hennawi {et~al.}(2006)Hennawi, Strauss, Oguri, Inada, Richards,
  Pindor, Schneider, Becker, Gregg, Hall, Johnston, Fan, Burles, Schlegel,
  Gunn, Lupton, Bahcall, Brunner, \& Brinkmann}]{Hennawi06}
---. 2006, \aj, 131, 1

\bibitem[{Hopkins {et~al.}(2010)Hopkins, Bundy, Croton, Hernquist, Keres,
  Khochfar, Stewart, Wetzel, \& Younger}]{Hopkins10b}
Hopkins, P.~F., {et~al.} 2010, \apj, 715, 202

\bibitem[{Kampczyk {et~al.}(2007)Kampczyk, Lilly, Carollo, Scarlata, Feldmann,
  Koekemoer, Leauthaud, Sargent, Taniguchi, \& Capak}]{Kampczyk07}
Kampczyk, P., {et~al.} 2007, \apjs, 172, 329

\bibitem[{Kaspi {et~al.}(2005)Kaspi, Maoz, Netzer, Peterson, Vestergaard, \&
  Jannuzi}]{Kaspi05}
Kaspi, S., Maoz, D., Netzer, H., Peterson, B.~M., Vestergaard, M., \& Jannuzi,
  B.~T. 2005, \apj, 629, 61

\bibitem[{Kellermann {et~al.}(1989)Kellermann, Sramek, Schmidt, Shaffer, \&
  Green}]{Kellermann89}
Kellermann, K.~I., Sramek, R., Schmidt, M., Shaffer, D.~B., \& Green, R. 1989,
  \aj, 98, 1195

\bibitem[{Kewley {et~al.}(2010)Kewley, Rupke, Jabran~Zahid, Geller, \&
  Barton}]{Kewley10}
Kewley, L.~J., Rupke, D., Jabran~Zahid, H., Geller, M.~J., \& Barton, E.~J.
  2010, \apj, 721, L48

\bibitem[{Kochanek {et~al.}(1999)Kochanek, Falco, \& Mu{\~n}oz}]{Kochanek99}
Kochanek, C.~S., Falco, E.~E., \& Mu{\~n}oz, J.~A. 1999, \apj, 510, 590

\bibitem[{Komossa {et~al.}(2003)Komossa, Burwitz, Hasinger, Predehl, Kaastra,
  \& Ikebe}]{Komossa03}
Komossa, S., Burwitz, V., Hasinger, G., Predehl, P., Kaastra, J.~S., \& Ikebe,
  Y. 2003, \apj, 582, L15

\bibitem[{Kormendy \& Richstone(1995)}]{Kormendy95}
Kormendy, J., \& Richstone, D. 1995, \araa, 33, 581

\bibitem[{Larkin {et~al.}(2006)Larkin, Barczys, Krabbe, Adkins, Aliado, Amico,
  Brims, Campbell, Canfield, Gasaway, Honey, Iserlohe, Johnson, Kress,
  LaFreniere, Lyke, Magnone, Magnone, McElwain, Moon, Quirrenbach, Skulason,
  Song, Spencer, Weiss, \& Wright}]{Larkin06}
Larkin, J., {et~al.} 2006, in SPIE Conference Series, Vol. 6269, 42

\bibitem[{Liu {et~al.}(2010{\natexlab{a}})Liu, Greene, Shen, \&
  Strauss}]{Liu10b}
Liu, X., Greene, J.~E., Shen, Y., \& Strauss, M.~A. 2010{\natexlab{a}}, \apj,
  715, L30

\bibitem[{Liu {et~al.}(2010{\natexlab{b}})Liu, Shen, Strauss, \&
  Greene}]{Liu10a}
Liu, X., Shen, Y., Strauss, M.~A., \& Greene, J.~E. 2010{\natexlab{b}}, \apj,
  708, 427

\bibitem[{Madau \& Quataert(2004)}]{Madau04}
Madau, P., \& Quataert, E. 2004, \apj, 606, L17

\bibitem[{Milosavljevi{\'c} \& Merritt(2001)}]{Milosavljevic01}
Milosavljevi{\'c}, M., \& Merritt, D. 2001, \apj, 563, 34

\bibitem[{Myers {et~al.}(2007)Myers, Brunner, Richards, Nichol, Schneider, \&
  Bahcall}]{Myers07}
Myers, A.~D., Brunner, R.~J., Richards, G.~T., Nichol, R.~C., Schneider, D.~P.,
  \& Bahcall, N.~A. 2007, \apj, 658, 99

\bibitem[{Myers {et~al.}(2008)Myers, Richards, Brunner, Schneider, Strand,
  Hall, Blomquist, \& York}]{Myers08}
Myers, A.~D., Richards, G.~T., Brunner, R.~J., Schneider, D.~P., Strand, N.~E.,
  Hall, P.~B., Blomquist, J.~A., \& York, D.~G. 2008, \apj, 678, 635

\bibitem[{Peng {et~al.}(2010)Peng, Ho, Impey, \& Rix}]{Peng10}
Peng, C.~Y., Ho, L.~C., Impey, C.~D., \& Rix, H.-W. 2010, \aj, 139, 2097

\bibitem[{Peterson {et~al.}(1998)Peterson, Wanders, Bertram, Hunley, Pogge, \&
  Wagner}]{Peterson98}
Peterson, B.~M., Wanders, I., Bertram, R., Hunley, J.~F., Pogge, R.~W., \&
  Wagner, R.~M. 1998, \apj, 501, 82

\bibitem[{Richstone {et~al.}(1998)Richstone, Ajhar, Bender, Bower, Dressler,
  Faber, Filippenko, Gebhardt, Green, Ho, Kormendy, Lauer, Magorrian, \&
  Tremaine}]{Richstone98}
Richstone, D., {et~al.} 1998, Nature, 395, 14

\bibitem[{Rodriguez {et~al.}(2009)Rodriguez, Taylor, Zavala, Pihlstr{\"o}m, \&
  Peck}]{Rodriguez09}
Rodriguez, C., Taylor, G.~B., Zavala, R.~T., Pihlstr{\"o}m, Y.~M., \& Peck,
  A.~B. 2009, \apj, 697, 37

\bibitem[{Rosario {et~al.}(2010)Rosario, Shields, Taylor, Salviander, \&
  Smith}]{Rosario10}
Rosario, D.~J., Shields, G.~A., Taylor, G.~B., Salviander, S., \& Smith, K.~L.
  2010, \apj, 716, 131

\bibitem[{Schneider {et~al.}(2007)Schneider, Hall, Richards, Strauss,
  Vanden~Berk, Anderson, Brandt, Fan, Jester, Gray, Gunn, SubbaRao, Thakar,
  Stoughton, Szalay, Yanny, York, Bahcall, Barentine, Blanton, Brewington,
  Brinkmann, Brunner, Castander, Csabai, Frieman, Fukugita, Harvanek, Hogg,
  Ivezic, Kent, Kleinman, Knapp, Kron, Krzesinski, Long, Lupton, Nitta, Pier,
  Saxe, Shen, Snedden, Weinberg, \& Wu}]{Schneider07}
Schneider, D.~P., {et~al.} 2007, \aj, 134, 102

\bibitem[{Scoville {et~al.}(2007)Scoville, Abraham, Aussel, Barnes, Benson,
  Blain, Calzetti, Comastri, Capak, Carilli, Carlstrom, Carollo, Colbert,
  Daddi, Ellis, Elvis, Ewald, Fall, Franceschini, Giavalisco, Green, Griffiths,
  Guzzo, Hasinger, Impey, Kneib, Koda, Koekemoer, Lefevre, Lilly, Liu,
  McCracken, Massey, Mellier, Miyazaki, Mobasher, Mould, Norman, Refregier,
  Renzini, Rhodes, Rich, Sanders, Schiminovich, Schinnerer, Scodeggio, Sheth,
  Shopbell, Taniguchi, Tyson, Urry, Van~Waerbeke, Vettolani, White, \&
  Yan}]{Scoville07b}
Scoville, N., {et~al.} 2007, \apjs, 172, 38

\bibitem[{Shen {et~al.}(2008)Shen, Vanden~Berk, Schneider, \& Hall}]{JJShen08}
Shen, J., Vanden~Berk, D.~E., Schneider, D.~P., \& Hall, P.~B. 2008, \aj, 135,
  928

\bibitem[{Shen {et~al.}(2010)Shen, Hall, Richards, Schneider, Strauss, Snedden,
  Bizyaev, Brewington, Malanushenko, Malanushenko, Oravetz, Pan, \&
  Simmons}]{Shen10}
Shen, Y., {et~al.} 2010, \apjs, submitted, arXiv:1006.5178

\bibitem[{Smith {et~al.}(2010)Smith, Shields, Bonning, McMullen, Rosario, \&
  Salviander}]{Smith10}
Smith, K.~L., Shields, G.~A., Bonning, E.~W., McMullen, C.~C., Rosario, D.~J.,
  \& Salviander, S. 2010, \apj, 716, 866

\bibitem[{Stockton {et~al.}(1998)Stockton, Canalizo, \& Close}]{Stockton98}
Stockton, A., Canalizo, G., \& Close, L.~M. 1998, \apj, 500, L121

\bibitem[{Stockton {et~al.}(2007)Stockton, Canalizo, Fu, \& Keel}]{Stockton07}
Stockton, A., Canalizo, G., Fu, H., \& Keel, W. 2007, \apj, 659, 195

\bibitem[{Valtonen {et~al.}(2008)Valtonen, Lehto, Nilsson, Heidt, Takalo,
  Sillanp{\"a}{\"a}, Villforth, Kidger, Poyner, Pursimo, Zola, Wu, Zhou,
  Sadakane, Drozdz, Koziel, Marchev, Ogloza, Porowski, Siwak, Stachowski,
  Winiarski, Hentunen, Nissinen, Liakos, \& Dogru}]{Valtonen08}
Valtonen, M.~J., {et~al.} 2008, Nature, 452, 851

\bibitem[{van Dam {et~al.}(2006)van Dam, Bouchez, Le~Mignant, Johansson,
  Wizinowich, Campbell, Chin, Hartman, Lafon, Stomski, \& Summers}]{Dam06}
van Dam, M.~A., {et~al.} 2006, \pasp, 118, 310

\bibitem[{Vestergaard(2002)}]{Vestergaard02}
Vestergaard, M. 2002, \apj, 571, 733

\bibitem[{Vestergaard \& Peterson(2006)}]{Vestergaard06}
Vestergaard, M., \& Peterson, B.~M. 2006, \apj, 641, 689

\bibitem[{Wainscoat \& Cowie(1992)}]{Wainscoat92}
Wainscoat, R.~J., \& Cowie, L.~L. 1992, \aj, 103, 332

\bibitem[{Wang {et~al.}(2009)Wang, Chen, Hu, Mao, Zhang, \& Bian}]{Wang09}
Wang, J.-M., Chen, Y.-M., Hu, C., Mao, W.-M., Zhang, S., \& Bian, W.-H. 2009,
  \apj, 705, L76

\bibitem[{Wizinowich {et~al.}(2006)Wizinowich, Le~Mignant, Bouchez, Campbell,
  Chin, Contos, van Dam, Hartman, Johansson, Lafon, Lewis, Stomski, Summers,
  Brown, Danforth, Max, \& Pennington}]{Wizinowich06}
Wizinowich, P.~L., {et~al.} 2006, \pasp, 118, 297

\bibitem[{Woo {et~al.}(2010)Woo, Treu, Barth, Wright, Walsh, Bentz, Martini,
  Bennert, Canalizo, Filippenko, Gates, Greene, Li, Malkan, Stern, \&
  Minezaki}]{Woo10}
Woo, J.-H., {et~al.} 2010, \apj, 716, 269

\bibitem[{Xu \& Komossa(2009)}]{Xu09}
Xu, D., \& Komossa, S. 2009, \apj, 705, L20

\end{thebibliography}

\begin{deluxetable}{cccccccccccccc} 
\rotate
\tabletypesize{\footnotesize}
\setlength{\tabcolsep}{.11cm}
\tablewidth{0pt}
\tablecaption{Double-peaked [O\,{\sc iii}] AGNs with Keck/LGSAO Images
\label{tab:sample}}
\tablehead{ 
\colhead{SDSS Name} & \colhead{$z$} & \colhead{$\Delta\theta$} &
\colhead{$\Delta S$} & \colhead{$K'_1$} & \colhead{$K'_2$} & \colhead{$\Delta V$} & 
\colhead{$L_{\rm [O III]}^b$} & \colhead{$L_{\rm [O III]}^r$} & 
\colhead{$L_{\rm 5GHz}$} & \colhead{$R$} &
\colhead{$\sigma_{\star}$} & \colhead{$M_{\rm BH}$} & \colhead{Ref} \\
 & & \arcsec\ & kpc & & & \kms\ & log(\lsun) & log(\lsun) & 
 log(W/Hz) & & \kms\ & log(\msun) & \\
\colhead{(1)} & \colhead{(2)} & \colhead{(3)} & \colhead{(4)} & 
\colhead{(5)} & \colhead{(6)} & \colhead{(7)} & \colhead{(8)} &
\colhead{(9)} & \colhead{(10)} & \colhead{(11)} & \colhead{(12)} &
\colhead{(13)} & \colhead{(14)} 
}
\startdata
\multicolumn{14}{c}{Merging Type-1 AGNs}   \nl
095207.6$+$255257&0.339&0.99& 4.8&15.3&16.9& 442&8.55&8.41& \nd& \nd&\nd& 8.4&3    \\
115523.7$+$150757&0.287&0.57& 2.5&14.3&15.4& 480&8.91&8.50&23.3&   2&\nd& 7.8&3    \\
123915.4$+$531415&0.202&1.24& 4.1&14.4&14.8& 715&7.87&7.76&24.1&  75&431$\pm$30& 7.8&3    \\
124859.7$-$025731&0.487&0.51& 3.1&15.5&16.7& 380&8.58&9.04& \nd& \nd&\nd& 8.8&3    \\
130724.1$+$460401&0.352&2.29&11.4&16.2&16.4& 518&8.30&8.32& \nd& \nd&293$\pm$57& 7.7&3    \\
150243.1$+$111557&0.390&1.39& 7.4&16.0&16.1& 657&9.84&8.96&24.3&  36&\nd& 9.4&3    \\
151735.2$+$214533&0.399&1.09& 5.9&15.5&18.1& 324&8.62&8.06& \nd& \nd&\nd& 7.5&3    \\
154107.8$+$203609&0.508&1.95&12.0&16.4&20.5& 441&8.96&8.63& \nd& \nd&\nd& 8.7&3    \\
\multicolumn{14}{c}{Merging Type-2 AGNs} \nl
110851.0$+$065901&0.182&0.64& 2.0&15.1&15.3& 204&8.74&8.27&23.6&  28&222& \nd&1    \\
114642.5$+$511030&0.130&2.71& 6.3&15.0&15.7& 287&8.03&8.17& \nd& \nd&175& \nd&1    \\
115715.0$+$081632&0.201&2.57& 8.5&14.5&14.7& 375&7.50&7.74& \nd& \nd&\nd& \nd&3    \\
124037.8$+$353437&0.161&0.22& 0.6&15.5&15.7& 491&8.79&8.20&23.2&  28&201& \nd&1    \\
133226.3$+$060627&0.207&1.54& 5.2&15.6&15.8& 409&7.79&8.04& \nd& \nd&275& \nd&1,3  \\
135646.1$+$102609&0.123&1.32& 2.9&14.7&15.0& 383&8.85&9.19&24.0& 128&251& \nd&1    \\
165206.1$+$310708&0.075&2.93& 4.2&14.3&20.6& 307&6.59&6.73& \nd& \nd&140& \nd&1,2  \\
172049.2$+$310646&0.095&0.40& 0.7&14.5&20.3& 389&7.69&7.29& \nd& \nd&242& \nd&1    \\
\multicolumn{14}{c}{Isolated Type-1 AGNs} \nl
120240.7$+$263139&0.476& \nd& \nd&14.9& \nd& 466&9.54&8.59&25.3&  66&\nd& 9.6&3    \\
121911.2$+$042906&0.555& \nd& \nd&15.2& \nd& 459&9.46&9.33& \nd& \nd&\nd& 8.5&3    \\
144012.8$+$615633&0.275& \nd& \nd&14.2& \nd& 505&8.89&8.42&23.4&   2&\nd& 7.7&3    \\
150437.7$+$541150&0.305& \nd& \nd&14.7& \nd& 719&9.02&8.21& \nd& \nd&\nd& 8.5&3    \\
153231.8$+$420343&0.209& \nd& \nd&15.1& \nd& 353&8.51&8.44& \nd& \nd&184$\pm$44& 8.2&3    \\
161027.4$+$130807&0.229& \nd& \nd&15.3& \nd& 326&7.94&8.11& \nd& \nd&175$\pm$19& 8.3&3    \\
161826.9$+$081951&0.446& \nd& \nd&14.3& \nd& 384&8.62&9.54&25.6& 137&\nd&10.3&3    \\
161847.9$+$215925&0.334& \nd& \nd&15.1& \nd& 365&8.84&8.49&23.8&   5&\nd& 9.0&3    \\
171930.6$+$293413&0.180& \nd& \nd&14.7& \nd&1105&8.12&7.67&22.7&   2&\nd& 7.4&3    \\
\multicolumn{14}{c}{Isolated Type-2 AGNs} \nl
040001.6$-$065254&0.171& \nd& \nd&14.1& \nd& 391&8.43&8.73&23.2&  10&242& \nd&1    \\
115249.3$+$190300&0.097& \nd& \nd&14.3& \nd& 323&7.61&7.55&22.4&   9&141& \nd&1,2  \\
120320.7$+$131931&0.058& \nd& \nd&13.6& \nd& 292&6.56&6.57&23.6& 156&\nd& \nd&2    \\
121957.5$+$190003&0.117& \nd& \nd&15.3& \nd& 390&7.74&7.44& \nd& \nd&123& \nd&1    \\
131106.7$+$195234&0.156& \nd& \nd&15.2& \nd& 314&7.27&7.35& \nd& \nd&150& \nd&1    \\
131236.0$+$500416&0.116& \nd& \nd&14.3& \nd& 353&7.11&7.01&22.7&  16&\nd& \nd&2    \\
132104.6$-$001446&0.082& \nd& \nd&14.8& \nd& 330&6.44&6.63& \nd& \nd&163& \nd&1    \\
140231.6$+$021546&0.180& \nd& \nd&14.4& \nd& 276&8.28&8.42&25.6&2553&222& \nd&1    \\
140534.8$+$244735&0.130& \nd& \nd&14.7& \nd& 218&8.19&7.60&22.2&   3&115& \nd&1    \\
140845.7$+$353218&0.166& \nd& \nd&14.1& \nd& 371&8.42&8.32&23.4&  16&250& \nd&1    \\
144157.2$+$094859&0.220& \nd& \nd&14.9& \nd& 826&8.73&8.24&23.7&  61&\nd& \nd&3    \\
145156.8$+$301603&0.158& \nd& \nd&14.6& \nd& 512&7.27&6.98& \nd& \nd&215& \nd&1    \\
150452.3$+$321415&0.113& \nd& \nd&14.3& \nd& 250&7.22&7.36&22.4&   7&\nd& \nd&2    \\
151659.2$+$051752&0.051& \nd& \nd&13.3& \nd& 326&7.05&7.06&22.2&   8&213& \nd&1,2  \\
151757.4$+$114453&0.227& \nd& \nd&14.7& \nd& 433&8.09&7.83& \nd& \nd&\nd& \nd&3    \\
155009.6$+$080839&0.232& \nd& \nd&14.7& \nd& 400&8.14&8.15&23.5&  23&259& \nd&1    \\
160027.8$+$083743&0.227& \nd& \nd&15.1& \nd& 422&7.80&7.83&25.5&3765&229& \nd&1    \\
160436.2$+$500958&0.146& \nd& \nd&14.4& \nd& 366&7.72&7.71& \nd& \nd&211& \nd&1,2  \\
160631.4$+$273643&0.158& \nd& \nd&15.4& \nd& 319&8.82&8.67&22.7&   9&121& \nd&1    \\
163056.8$+$164957&0.034& \nd& \nd&13.5& \nd& 301&7.25&7.45&21.4&   2&107& \nd&1    \\
163316.0$+$262716&0.071& \nd& \nd&14.4& \nd& 307&6.70&6.50& \nd& \nd&\nd& \nd&2    \\
171544.0$+$600835&0.157& \nd& \nd&14.8& \nd& 348&8.54&8.36&23.6&  62&139& \nd&1,3  \\
230442.8$-$093345&0.032& \nd& \nd&12.1& \nd& 320&6.46&6.35&22.0&   8&134& \nd&1,2  \\
233313.2$+$004912&0.170& \nd& \nd&13.8& \nd& 434&8.29&8.67&25.0&1234& 98& \nd&1    \\
235256.6$+$001155&0.167& \nd& \nd&14.8& \nd& 322&7.98&7.86& \nd& \nd&219& \nd&1    
\enddata
\tablecomments{
Double-peaked \OIII\ AGNs with Keck/LGSAO images.
Column 1: J2000 designation.
Column 2: Redshift.
Columns 3,4: Projected angular separation (arcsec) and physical
separation (kpc) between the main components in a merging system.
Columns 5,6: For merging systems, these are the $K'$ magnitudes of the
main components; for isolated systems, $K'_1$ lists the $K'$
magnitude of the source, while $K'_2$ is left blank.
Column 7: Velocity splitting (\kms) between the \OIII\,$\lambda 5007$
components: $\Delta V/c = [(1+z_r)^2/(1+z_b)^2-1]/[(1+z_r)^2/(1+z_b)^2+1]$, where $z_r$ and $z_b$ are the redshifts of the redshifted and blueshifted \OIII\ components, respectively. Typical statistical errors are 16 \kms.
Columns 8,9: \OIII\,$\lambda 5007$ luminosity in log(\lsun) for the
blueshifted ($b$) and redshifted ($r$) line, corrected for Galactic extinction.
Column 10: Radio luminosity at rest-frame 5 GHz in log(W/Hz), {\it K}-corrected from the observed 1.4 GHz fluxes assuming $L_\nu \propto \nu^{-0.7}$. 
Column 11: \citet{Kellermann89} radio loudness, $R = F_{\nu,\rm 5GHz}/F_{\nu,\rm 4400A}$.
Column 12: Stellar velocity dispersion (\kms), corrected for
3\arcsec\ fiber size. For type-1 sources, we list our own measurements; for type-2 sources, these are compiled from \citet{Liu10a}.
Column 13: Black hole mass in log(\msun), estimated using the H$\beta$ calibration of \citet{Vestergaard06}. The errors are about 0.5 dex.
Column 14: Source references---(1) \citet{Liu10a}, (2) \citet{Wang09}, (3) \citet{Smith10}.
}
\end{deluxetable}

\end{document}